\documentclass[aps,twocolumn,showpacs]{revtex4}
\usepackage{graphics}
\begin{document}
\tighten
\bibliographystyle{apsrev}
\def\half{{1\over 2}}
\def \D {\mbox{D}}
\def\curl {\mbox{curl}\,}
\def \ep {\varepsilon}
\def \lleq {\lower0.9ex\hbox{ $\buildrel < \over \sim$} ~}
\def \ggeq {\lower0.9ex\hbox{ $\buildrel > \over \sim$} ~}
\def\beq{\begin{equation}}
\def\eeq{\end{equation}}
\def\ber{\begin{eqnarray}}
\def\eer{\end{eqnarray}}
\def \apl {ApJ, }
\def \aps {ApJS, }
\def \pd {Phys. Rev. D, }
\def \prl {Phys. Rev. Lett., }
\def \pl {Phys. Lett., }
\def \np {Nucl. Phys., }
\def \l {\Lambda}

%%\textheight 21.0cm
%%\textwidth 16cm
%%\leftmargin -4cm
%%\topmargin 0cm
%%\topmargin -2cm
%\begin{document}
%\draft
\title{Aspects of Tachyonic Inflation with Exponential Potential}
\author{M. Sami}
\email{sami@jamia.net}
\author{Pravabati Chingangbam}
\email{prava@jamia.net}
\author{Tabish Qureshi}
\email{tabish@jamia.net}
\affiliation{Department of Physics, Jamia Millia Islamia,
 New Delhi-110025, INDIA}

\begin{abstract}
We consider issues related to tachyonic inflation with exponential potential.
We find  exact solution of evolution equations in the slow roll limit in FRW
cosmology. We also carry out similar analysis in case of Brane assisted
tachyonic inflation. We investigate the phase space  behavior of the system
and show that the dust like solution is a late time attractor. The difficulties
associated with reheating in the tachyonic model are also indicated.
 \end{abstract}
\maketitle 

\section{INTRODUCTION}
M/String theory inspired models are under active consideration in cosmology
at present. It has recently been suggested that rolling tachyon condensate, in
a class of string theories, may have interesting cosmological consequences.
Sen\cite{sen1,sen2} has shown that the decay of  D-branes produces a pressureless gas 
with finite energy density that resembles classical dust. Rolling tachyon
matter associated with unstable D-branes has an interesting equation of state
which smoothly interpolates between -1 and 0. As the tachyon field rolls
down the hill, the universe undergoes accelerated expansion and at a
particular epoch   the scale factor passes through the point of inflection
marking the end of inflation. At late times the energy  density of tachyon
matter scales as $a^{-3}$. The tachyonic matter, therefore, might provide an
explanation for inflation at the early epochs and could contribute to some
new form of cosmological dark matter at late 
times \cite{gibbons,fairbairn,feinstein,paddy1,shiu,paddy2,jatkar,hashimoto,
kim,shima,minahan,cornalba,benoum,li}.
The evolution of
FRW cosmological models and linear perturbations of tachyon matter rolling
towards a minimum of its potential is discussed by  Frolov, Kofman and Starobinsky \cite{star}.
 Effective
potential for tachyon field is computed in reference \cite{moore}; the
expression is exact in $\alpha'$ but tree level in $g_s$. Sen \cite{sen3} has shown
that the choice of an exponential potential for the tachyonic field leads to the
absence of plane-wave solutions around the tachyon vacuum and
exponential decay of the pressure at late times. A homogeneous tachyon
field evolves towards its ground state without oscillating about it. Therefore,
the  conventional reheating mechanism in tachyonic model does not work.
Quantum mechanical particle production during inflation  provide an alternative
mechanism by means of which the universe could reheat. However, the
energy density of radiation  so created is much smaller than the energy
density of the field. In case of non-tachyonic field, this led to the requirement 
of steep inflaton potential so that the field energy density could decay faster
than the radiation energy density and  radiation domination could commence.
Brane  assisted inflation was then invoked to support inflation with steep
potentials. A detailed account of reheating via gravitational particle
production in brane world scenario is discussed in reference \cite{copeland,sahni}.  

In this paper we discuss issues of inflation with  tachyon rolling down an
exponential potential. We  find exact solution of slow roll equations in usual
4-dimensional FRW cosmology as well as on the brane. We carry out the
phase space analysis for the system under consideration and find out the
fixed point and discuss its stability. We also indicate the problems of reheating
in the tachyonic model. Effects of tachyons in the context of brane world cosmology
are discussed in reference\cite{btachyon}. Dynamics of gauge fields with rolling tachyon on unstable D-branes
is studied in reference \cite{minahan2,akira,mehen}.

As recently demonstrated by Sen \cite{sen1,sen2}
 a rolling
tachyon condensate in a spatially flat FRW cosmological model is described
by an effective fluid with energy momentum tensor
$T^{\mu}_{\nu}=diag\left(-\rho,p,p,p\right)$, where the energy density $\rho$
and pressure p are given by
\begin{equation}
\rho={V(\phi) \over {\sqrt{1-\dot{\phi}^2}}}
\end{equation}
\begin{equation}
p=-V(\phi)\sqrt{1-\dot{\phi}^2}
\end{equation}
The Friedmann equation takes the form 
\begin{equation}
H^2={1 \over 3M_p^2} \rho \equiv {1 \over 3M_p^2}{V(\phi) \over {\sqrt{1-\dot{\phi}^2}}}
\end{equation} 
The equation of motion of the  tachyon field minimally coupled to gravity   is 
\begin{equation} 
{\ddot{\phi} \over {1-\dot{\phi}^2}}+3H \dot{\phi}+{V_{,\phi} \over V({\phi})}=0
\end{equation}
The conservation equation equivalent to (4) has the usual form
$$ {\dot{\rho}_{\phi} \over \rho_{\phi}}+3H (1+\omega)=0 $$
where $\omega \equiv {p_{\phi} \over \rho_{\phi}}= \dot{\phi}^2-1$ is the
equation of state for the tachyon field. Thus a universe dominated by tachyon
field would go under accelerated expansion as long as $\dot{\phi}^2\ <\ {2
\over 3}$  which is very different from the condition of inflation for non-tachyonic
field, $\dot{\phi}^2\ <\ V(\phi)$.
It is obvious that the tachyon field should starts rolling with a small value of
$\dot{\phi}$ in order to have a long period of inflation.

\subsection{DYNAMICS OF TACHYONIC INFLATION IN FRW COSMOLOGY}
The exponential potential 
$$ V(\phi)=V_0e^{-\alpha \phi}  $$
has played an important role within the inflationary cosmology. Sen 
\cite{sen3} has recently argued in support of this potential for the 
tachyonic system. 
The field equations (3) and (4) for tachyonic matter with the exponential potential can be solved exactly in the
slow roll limit. In this limit 
\beq
{\dot{a}(t) \over a(t)} \simeq \sqrt{{1 \over 3M_p^2} V(\phi)},
\eeq
and (4) becomes
\beq
3 H \dot{\phi}\simeq -{V_{,\phi} \over V(\phi)} = \alpha
\eeq
Substitution of (5) in (6) leads to
\beq
\dot{\phi}={ \alpha \over {3\beta}}e^{\alpha \phi \over 2}
\eeq
where $\beta=\sqrt{ V_0/3M_p^2}$. Equation (7) immediately integrates to yield
\beq
\phi(t)=-{2 \over \alpha}\ln \left[C-{\alpha^2 \over {6 \beta}} t\right]
\eeq
where $ C=e^{-\alpha \phi_i \over 2}$ \ and $\phi(t=t_i=0)\equiv \phi_i$. 
Putting this expression for $\phi(t)$ in (5) we get
\beq
{a(t) \over a_i}= e^{\beta t\left(C-{\alpha^2 \over {12 \beta}}t \right)}
\eeq
This equation tells us that the scale factor passes 
through an inflection point at
$$t=t_{end}={6 \beta \over {\alpha^2}}\left(C-{\alpha \over {\sqrt6 \beta}} 
                                     \right)$$
Thus $t_{end}$ marks the end of inflation. From Eq. (8) we find
\beq
\dot{\phi}_{\it end}=\sqrt{2 \over 3},  \  
\phi_{end}=-{1 \over \alpha} \ln \left( {\alpha^2 \over {6\beta^2}} \right), \  
V_{end}={{\alpha^2 M_p^2} \over 2}
\eeq
The expression for the number of inflationary e-foldings can easily be established
\beq
{\cal{ N}} = \ln{{a(t)} \over{ a_i}}= \int_{t_i=0}^t { H( t') dt'}=\beta t \left(C-{\alpha^2 \over {12 \beta}}t \right)
\eeq
Using Eqs. (10) and (11) we find
\beq
V_{end}={Vi \over {{2\cal N}+1}}
\eeq
It is interesting to note that the expression (12) is very similar to what one gets
in case of a normal scalar field with exponential potential  propagating on the
Brane. As the condition $\dot{\phi}^2 <\ 2/3\ $ is very different from the
condition of inflation for a normal scalar field, the slow roll parameters
$\epsilon$ and $\eta$  assume an unusual form in the case of tachyonic field, for 
instance 
\beq
\epsilon={M_p^2 \over 2} \left({V_{,\phi} \over V}\right)^2{1 \over V}
\eeq
which resembles the slow roll parameter in brane world cosmology \cite{cline}. As noted
by Fairbairn and Tytgat \cite{fairbairn} that for small $\dot{\phi}$ near the top of the potential we can
use the canonically normalized field $\phi_c=\sqrt V_0 \phi$ which brings the parameters
to their usual form \cite{Liddle}. We will also use this crude approximation to estimate
the magnitude of perturbations.The detailed description of perturbations is given by Hwang and Noh \cite{hwang}.The COBE normalized value for the amplitude of scalar 
density perturbations
\beq
\delta_H^2\simeq {1 \over {75 \pi^2}}{V_i^2 \over {\alpha^2 M_p^2}}\simeq 4\times  10^{-10}
\eeq
 can be used to estimate $ V_{end}$ as well as $ \alpha$. Here $ V_i$ refers to the value of the 
 potential at the commencement of inflation. Using Eqs. (12) and (14) 
with ${\cal N} =60$ we obtain
 \beq
 V_{end}\simeq 4 \times 10^{-11}M_p^4
 \eeq
 At the end of inflation, apart from the field energy density, a small amount of radiation is also 
 present due to particles being produced quantum mechanically during inflation \cite{fordfos}
 \beq
 \rho_r = 0.01 \times g_p H_{end}^4 ~~~~~~~(10 \le g_p \le 100)
 \eeq
 which shows that the field energy density far exceeds the density in the radiation
 \beq
 {\rho_r \over{ \rho_{\phi}}} \simeq 0.01\times g_p {V_{end} \over { 9 M_p^2}}\simeq 4\times g_p \times 10^{-14}
 \eeq 
 
\subsection{TACHYONIC INFLATION ON THE BRANE}

The prospects of inflation in Brane World scenario \cite{randall}
improve due to the presence of an additional quadratic density term in the
Friedmann equation\cite{cline,copeland}. As a result, the class of steep potentials can
successfully describe inflation on the brane. In the 4+1 dimensional brane
world scenario inspired by Randall-Sundrum model\cite{randall}, the Friedmann
equation is modified to\cite{cline}
\begin{equation} H^2={1 \over 3M_p^2} \rho_{\phi} \left(1+{\rho_{\phi} \over 2\lambda_b}
\right)+ {\Lambda_4 \over 3}+{ {\cal E} \over a^4} \end{equation} where $\cal E$ is
an integration constant which transmits bulk graviton influence onto
the brane and $\lambda_b$ is the brane tension.  For simplicity we set
$\Lambda_4$ equal to zero and also drop the last term as 
inflation would render it so, leading to the expression 
\begin{equation}
H^2={1 \over 3M_p^2} \rho_{\phi} \left(1+{\rho_{\phi} \over 2\lambda_b} \right)
\end{equation} 
where $ \rho_{\phi} $ is given by (1) if one is dealing with universe 
dominated by a single tachyon field minimally coupled to gravity. The brane
effects, in context of inflation, are most pronounced in the high energy limit
$V>> \lambda_b$ ; the Friedmann equation in this limit becomes 
\begin{equation} 
H={\rho_{\phi} \over{(6\lambda_b M_p^2)^{1/2}}} 
\end{equation}  
Analogous to the preceding section equations (6) and (20)
can be solved exactly to obtain
\beq
\phi(t)=-{1 \over \alpha}\ln \left[C-{\alpha^2 \over {3 \beta_b}} t\right]
\eeq
where $ C=e^{-\alpha \phi_i}\  $ and $\beta_b=\left(V_0^2/6 \lambda_bM_p^2\right)^{1/2}$.
The scale factor is given by 
\beq
{a(t) \over a_i}= e^{\beta_b t\left(C-{\alpha^2 \over {6 \beta_b}}t \right)}
\eeq
At the point of inflection 
$ t=t_{end}={3 \beta_b \over \alpha^2} \left(C-{\alpha \over {\sqrt3 \beta_b}}\right)$ we have
\beq
V_{end}= {V_0\alpha \over {\sqrt 3 \beta_b}}=\alpha M_p \sqrt{2 \lambda_b} \quad  \   \dot{\phi}_{end}= \sqrt {1 \over 3}.
\eeq
which is consistent with the expression of the slow roll parameter on the brane
$$\epsilon={M_p^2 } \left({V_{,\phi} \over V}\right)^2{2\lambda_b \over V^2}$$
The number of e-foldings is related to $ V_{end}$ and $V_i$ as
\beq
V_{end}=\left({V_i^2
 \over {{2\cal N}+1}} \right)^{1/2}
\eeq

Similar to the previous section the COBE normalized value for the 
amplitude of scalar density perturbations can be used to estimate both 
$V_{\it end}$ and $\lambda_b$. The ratio of the radiation density (created during
inflation) to the field energy density turns out again to be a small number
$${\rho_r \over \rho_{\phi}}\simeq g_p\times 10^{-16}  $$
independent of $ \alpha$.

\section{Phase portrait}

In this section we carry out phase space analysis\cite{belinsky} for  tachyonic 
field with exponential potential $V=V_0e^{-\alpha \phi}$ evolving 
in a spatially flat FRW 
universe. The tachyon field evolution equation and the Friedmann constraint 
equation can be cast in the form
\beq
x'=\left(1-x^2 \right)(1-3z x)
\eeq
\beq
y'=-x y
\eeq
\beq
z^2={y \over 3} { 1 \over \sqrt{(1-x^2)}}
\eeq
where we have introduced the dimensionless variables
$$  x  =  {\dot{ \phi}},\quad y={V(\phi) \over {\alpha^2 M_p^2}},
               \quad z={{H}\over{\alpha}}$$
and prime denotes derivative with respect to the dimensional 
variable $\eta=\alpha t $. To find out the critical points we set 
r.h.s of Eqs. (25) and (26) to zero and we obtain
 two fixed points $(\pm 1,0)$ of which $(-1,0)$ is not physical.
In order to study the stability of the critical point we perturb about
the critical point
\beq 
 x=1-u,\quad y=v
\eeq
where $u$ and $v$ are infinitesimally close to the critical point $(1,0)$.
Putting them in the coupled equations and keeping only linear terms we get
\beq 
u' \simeq -2u
\eeq
\beq
v' \simeq -v
\eeq
whose solutions are
\begin{eqnarray*}
u &=& u_0e^{-2\eta}\\
v &=& v_0e^{-\eta}
\end{eqnarray*}
which clearly demonstrates the stability of the critical point.
Asymptotically $\ {\dot\phi}(\infty)=1\ $  and $\ V(\infty)=0\ $  implying 
that $ a(t)\propto 
t^{2/3}$ and that pressure tends to zero keeping the energy density finite 
and non-zero \cite{paddy1}. We, 
therefore, conclude that dust like solution in the tachyonic model is a 
late time attractor.\

\begin{figure}
\resizebox{3.0in}{!}{\includegraphics{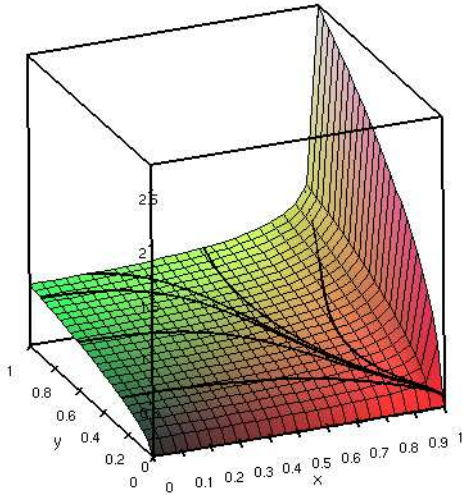}}
\caption{
Plot of the phase trajectories on the constraint surface. Trajectories 
starting anywhere in the phase space end up at the stable critical point (1,0).}
\end{figure}

\subsection{REHEATING PROBLEM IN THE TACHYONIC MODEL}
The tachyonic field could  play the dual role of inflaton at the early
epochs and of some new form of dark matter at late times. However, as
indicated by Kofman and Linde \cite{kofman}, reheating is problematic in these models. As
emphasized by Sen \cite{sen1,sen2} a homogeneous tachyon field evolves towards its
ground state without oscillating about it. Hence the  conventional reheating
mechanism in tachyonic model does not work. Quantum mechanical particle
production during inflation  provides an alternative mechanism by means of
which the universe could reheat. But the energy density of
radiation so created is much smaller than the energy density in the field. In
case of non-tachyonic field, this led to the requirement  of a steep inflaton
potential so that the field energy density could decay faster then the radiation
energy density and  radiation domination could commence. Unfortunately, from
the general arguments in case of the tachyonic matter, it seems that the field 
energy density after inflation, always scales slower than radiation. Hence
radiation domination will never commence in this model irrespective of the
steepness of the tachyonic potential. 
\begin{figure}[h]

\resizebox{!}{2.5in}{\includegraphics{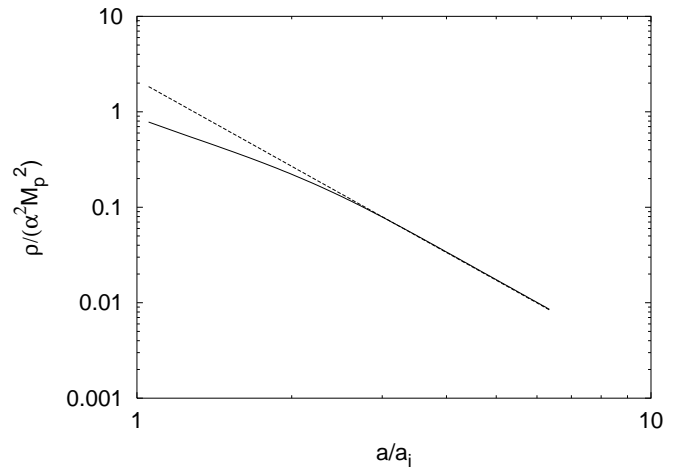}}
\caption{
The post inflation evolution of tachyonic field energy density (solid line) is shown as a function of the 
expansion factor. Immediately after inflation the field density scales as 
$\ a^{-2}$. The decay law $ \rho_{\phi}\propto a^{-3}$ is shown by the 
dashed line. It is seen clearly that the energy density in tachyonic 
field fast approaches $\ a^{-3}$.  }
\end{figure}

\noindent In fact, $\rho_{\phi} \propto a^{-3\dot{\phi}^2}$ for tachyonic field,
 as
$\omega=\dot{\phi}^2-1$ in this case . Thus immediately after the  inflation has ended $(
\dot{\phi}^2=2/3) $,
the field energy density $\rho_{\phi}$ scales as $a^{-2}$.
Gibbons has shown that $\dot{\phi}$ is a monotonically increasing function of time with
maximum value equal to one. Therefore, at best $\rho_{\phi}$ can scale as
$a^{-3}$. This argument is valid irrespective
of the form of tachyonic potential provided it belongs to the class of potentials such that
$V(\phi) \rightarrow 0$ as $ \phi \rightarrow \infty$.
 Hence, radiation energy density created at the end of inflation
would redshift faster then the energy density in the tachyonic field. We have
evolved the  tachyonic field equations for exponential potential numerically; 
Fig.2 displays the field energy density versus the scale factor and shows that
tachyonic matter comes to dominate very early after inflation has ended 
lending
support to the analysis of Kofman and Linde\cite{kofman}. 

To sum up, we have studied tachyonic inflation with exponential 
potential and found the exact solution of evolution equations in the 
slow roll limit in FRW cosmology as well as on the brane.
We have shown that the dust like solution is a late time attractor of 
the tachyonic system. We have also pointed out that reheating is 
problematic in the tachyonic model.

\begin{acknowledgements}
We are thankful to V. Sahni for useful discussions.
\end{acknowledgements}
%\section{References}
%\begin{references}

\end{document}